\documentclass[preprint,aps]{revtex4}
\usepackage{graphicx}
\usepackage{dcolumn}
\usepackage{bm}

\begin{document}

\title{\bf Raman spectroscopy of iodine-doped double-walled carbon nanotubes}

\author{J. Cambedouzou, and J.-L. Sauvajol}
\affiliation{Groupe de Dynamique des Phases Condens\'ees (UMR CNRS 
5581), Universit\'e Montpellier II, 34095 Montpellier Cedex 5, France}

\author{A. Rahmani}
\altaffiliation[Also at ]{Groupe de Dynamique des Phases Condens\'ees (UMR CNRS 
5581), Universit\'e Montpellier II, 34095 Montpellier Cedex 5, France} 
\affiliation{ D\'epartement de Physique, Universit\'e MY Ismail, Facult\'e des Sciences, BP 4010,  
50000 Mekn\`es, Morocco}

\author{E. Flahaut, A. Peigney and C. Laurent}
\affiliation{ Centre Interuniversitaire de Recherche et d'ing\'enierie des Mat\'eriaux, (UMR CNRS 
5085), Universit\'e Paul Sabatier, 31602 Toulouse Cedex 4, France}

\date{\today}

\vskip 1.5cm
\vskip 1cm

\begin{abstract}

We present a Raman spectroscopy study of iodine-intercalated (p-type doped) double-walled carbon nanotubes. Double-walled carbon nanotubes (DWCNTs) are synthesized by catalytic chemical vapor deposition (CCVD) and characterized by Raman spectroscopy. The assignment of the radial breathing modes and the tangential modes of pristine DWCNTs is done in the framework of the bond polarization theory, using the spectral moment method. The changes in the Raman spectrum upon iodine doping are analysed. Poly-iodine anions are identified, and the Raman spectra reveal that the charge transfer between iodine and DWCNTs only involves the outer tubes. 

\end{abstract}

\vskip 1.5cm

\maketitle

\section{Introduction}

Double-walled carbon nanotubes (DWCNTs) are in the border between single-walled carbon nanotubes (SWCNT) and multi-walled carbon nanotubes (MWCNT). Two main processes are used to synthesize DWCNT, namely the direct route and the "peapods" conversion route. The first route is the DWCNTs synthesis  by catalytic chemical vapor deposition (CCVD) using cobalt, nickel or iron as catalysts \cite{flahaut,zhu}. The second route is the coalescence of C$_{60}$ chains confined into SWNTs (the so-called "peapods") giving the inner tube of the DWCNT \cite{bandow}. DWCNT prepared from the conversion of "peapods" were found to be arranged in large bundles \cite{abe} with a small distribution of tube diameters. On the contrary, in DWCNTs prepared by the CVD method, the tubes remain mainly isolated. However, the tubes have a broader diameter distribution\cite{flahaut}. One of the applications of double-walled carbon nanotubes is the realization of nanocylindrical capacitors. Recently it was shown that such a capacitor can be obtained by the adsorption of bromine anions at the surface of the outer tubes of DWCNT \cite{chen}.\newline
Resonant Raman scattering is known as a very powerful tool for the investigation of pristine \cite{dresselhaus} and intercalated, and consequently doped, single-walled carbon nanotubes \cite{rao1,bendiab1,bendiab2,bendiab3}. More recently Raman studies on pristine \cite{bandow,bacsa} and doped \cite{chen} double-walled carbon nanotubes (DWCNT) were performed \cite{chen}. It was found that the Raman spectrum in the radial breathing modes (RBM) range gives information about the diameter distribution and on the changes in the strength of RBM upon doping. The tangential modes (TM) range of Raman spectra gives information about the metallic or semiconducting character of nanotubes and on the changes of these features induced by the charge transfer between the dopant and the tubes. \newline

In the present work, we investigate the changes in the Raman spectra upon iodine doping of double-walled carbon nanotubes prepared from a CCVD route. This study completes the recent Raman investigation concerning bromine doping of DWCNT obtained from "peapod" conversion \cite{chen}.

\section{Experimental section}

Pristine double-walled carbon nanotubes were synthesized by catalytic chemical vapor deposition (CCVD) of a mixture of CH$_{4}$ (18 mol. \%) in H$_{2}$ on a MgO-based catalyst, at a temperature of 1000$^\circ$C. The complete procedure of synthesis of DWCNTs is presented elsewhere \cite{flahaut,flahaut2}. Transmission electronic microscopy (TEM) observations of pristine DWCNT (figure 1.a) show that the carbon nanotubes are clean (no amorphous deposit) and generally isolated, or gathered into small-diameter bundles, mainly composed of DWCNTs \cite{flahaut2,colomer}. DWNTs represent more than 75 \% of the whole population of the sample. The presence of some larger well-organised DWNTs bundles has also been observed and the detailed study by electron diffraction has been reported elsewhere \cite{colomer}. The inner and outer diameters range from 0.5 to 2.5 nm and from 1.2 to 3.2 nm, respectively \cite{flahaut2}. Some examples of DWCNTs with different diameters are shown in figures 1.b-d. These experimental observations are in good agreement with the diameters which have been calculated from the frequency of the RBM. Other multi-walled nanotubes can be present in the samples. For instance, a triple-walled carbon nanotube is identified on figure 1.e. \newline
Before doping, DWCNT samples were annealed at 250$^\circ$C under high vacuum for about 65 hours in order to remove adsorbed gazes. Iodine ions doping of DWCNT samples was done following the procedure previously described for doping single-walled nanotube bundles to saturation \cite{grigorian}. \newline
Room-temperature Raman spectra were measured using the Ar/Kr laser lines at 514.5 nm (2.41 eV) and 647.1 nm (1.92 eV) in the back-scattering geometry on a triple substractive Jobin-Yvon T64000 spectrometer equipped with a liquid nitrogen cooled charge coupled device (CCD) detector. Room-temperature Raman spectra excited at 1064 nm (1.16 eV) were measured using a Bruker Fourier-transform spectrometer and a CCD detector.\newline

\section{Results and discussion}

Figure 2 (top) shows the RBM range and the TM range of the Raman spectrum of a pristine DWCNT sample measured using a 1064 nm excitation wavelength. RBMs at different wavenumbers are clearly visible in figure 2 (top-left), indicating nanotubes with different diameters in the sample. RBMs can be separated into two groups. The first group contains three main peaks at 150 cm$^{-1}$, 166 cm$^{-1}$, and 175 cm$^{-1}$, and the second group contains three peaks at 265 cm$^{-1}$, 313 cm$^{-1}$, and 330 cm$^{-1}$.\newline 
Calculations have been performed in order to find a simple relation between the RBM frequencies and the tube diameters of the double-walled carbon nanotube \cite{popov,bandow2,damnjanovic1}. Recently \cite{rahmani}, Raman responses in finite and infinite isolated DWCNT were calculated in the framework of the bond-polarization theory, using the spectral moment method \cite{rahmani2}. In all these calculations, DWCNT consist of two nanotubes, the walls of which are assumed to be at a distance close of 0.34 nm and the diameters of which are $D_{inner}$ and $D_{outer}$. The C-C intratube interactions are described by using the same force constants set that the one used in our calculation of the Raman spectrum of isolated single-wall carbon nanotubes \cite{rahmani2}. A Lennard-Jones potential, $U_{LJ}=4 \epsilon [(\sigma /R)^{12}-(\sigma /R)^{6} ]$, is used to describe the van der Waals intertube interactions between inner and outer tubes in DWCNT. The values of the Lennard-Jones parameters are kept fixed at $\epsilon $= 2.964  meV and $\sigma$ = 3.407 \AA \cite{rahmani2}. The main conclusions of these calculations are summerized below. For DWCNT of relative small diameters and then featured by a large gap between RBMs, each mode can be identified with respect to its dominant vibrational contribution providing from the inner or the outer tube respectively (see also figure 2 in Ref.\cite{popov}). Since RBMs of our DWCNT samples present large energy gap, in the following, we will assign, for simplicity, the different modes to specific inner or outer tubes. Our calculations \cite{rahmani} show that the frequencies of the RBM and TM of DWCNT significantly differ from those calculated for isolated single-walled carbon nanotubes (SWCNT). It was found for inner tubes with diameters in the range 0.6-2.2 nm, that the shift of the RBM in DWCNT with respect to its position as an isolated SWCNT depends linearly on the diameter. The diameter dependence of the frequency of the RBM ($A_{1}$ symmetry) assigned to the inner tube is well fitted by the phenomenologic expression:\\ 

\begin{equation}
\omega _{inner}(cm^{-1}) =\frac{A_i}{D_{inner}}+B_i D_{inner}+C_i
\end{equation}
\\
with parameters close to: $A_i$= 225 $cm^{-1}nm$, $B_i$=20.6 $cm^{-1}nm^{-1}$, $C_i$=-2.4 $cm^{-1}$ 
\\
For the corresponding outer tubes with a diameter range 1.2-3 nm, the diameter dependence of the RBM frequency assigned to the outer tube is fitted by:\\

\begin{equation}
\omega _{out}(cm^{-1}) =\frac{A_o}{D_{outer}^2}+ \frac{B_o}{D_{outer}}+C_o
\end{equation}
\\
with parameters close to: $A_o$=-88.76 $cm^{-1}nm^{2}$, $B_o$=324.1 $cm^{-1}nm$, $C_o$=-14.7 $cm^{-1}$\newline 
\\
The measurements of RBM frequencies permit to derive the inner and outer diameters from these expressions \cite{rahmani}.\\
Using these latter results, we assign each RBM of the second group to a tube with a specific inner diameter: 9.05 \AA (265 cm$^{-1}$), 7.50 \AA (313 cm$^{-1}$) and 7.05 \AA (330 cm$^{-1}$), respectively. Each RBM of the first group is assigned to an outer diameter of 15.9 \AA (150 cm$^{-1}$), 14.35 \AA (166 cm$^{-1}$) and 13.85 \AA (175 cm$^{-1}$), respectively. In a single-resonance approach, consistently with energy transitions calculation, incident and scattered photons can resonate with the $E^{sc}_{11}$ for the semi-conducting inner tubes and $E^{sc}_{22}$ for the semi-conducting outer tubes \cite{kataura}.  The calculated Raman response for a sample which contains DWCNT having the previous mentioned diameters (e.g. (9,4)@(18,4), (6,5)@(16,4) and (7,3)@(16,3)) is given in figure 2 (top-left, bottom line). The difference in the intensity between experimental and calculated spectra is mainly due to the resonant character of the Raman scattering which is not taken into account in these calculations. The tangential modes (TM) range is shown in figure 2 (top-right). In a first approximation, the TM profile can be fitted by using three Lorentzians. In agreement with recent calculations \cite{rahmani} the single Lorentzian at the highest frequency (1591 cm$^{-1}$) represents the unresolved $A_{1}$ and  $E_{1}$ tangential modes for the outer tubes, while the intermediate peak at 1584 cm$^{-1}$  is associated to the $A_{1}$ tangential mode for the inner tubes. A broad component around 1548 cm$^{-1}$ is also observed. It can possibly be assigned to a mixing of the $E$ modes of the inner tubes (the lowest frequency modes calculated in DWCNT \cite{rahmani}) and the $E$  mode of the outer tubes ($E_{1}$ or $E_{2}$ modes as function of the chirality angle \cite{rahmani2}). \newline
Figure 2 (bottom) shows the changes in the Raman spectrum induced by iodine doping. First we focus on the response in the RBM range. The comparison of figures 2, top-left and bottom-left, states that the main RBM between 230 cm$^{-1}$ and 400 cm$^{-1}$ are present with close intensities in the two spectra. Consequently, we claim that the iodine doping does not affect the RBMs attributed to the semiconducting inner tubes of DWCNTs.  In contrast, the intensity of the RBM around 165 cm$^{-1}$ and 176 cm$^{-1}$ is significantly reduced upon doping. However, the line at 152 cm$^{-1}$ seems to be stronger than in the pristine sample. It was found in iodine-doped multi-walled carbon nanotubes, that the frequency of $I_{5}^{-}$ species downshifts when the excitation energy decreases (from 170 cm$^{-1}$ at 514.5 nm to 158 cm$^{-1}$ at 633 nm \cite{zhou}). As a consequence we assign this line to be the signature of poly-iodine anions in the iodine-doped DWCNT sample. The main conclusion is that iodine doping only affects the RBM of the outer tubes, suggesting adsorption of iodine species only at the surface of the outer tubes.
The dependence of the tangential modes upon doping leads to a similar conclusion. In the doped sample, (i) the strongest peak is shifted by 6 cm$^{-1}$ to higher wavenumbers (from 1591 cm$^{-1}$ to 1597 cm$^{-1}$), (ii) in contrast, the component at 1586 cm$^{-1}$, assigned to the TM ($A_1$ symmetry) for the inner tubes, and the band at 1548 cm$^{-1}$ are unshifted. Furthermore, we observe a significant decrease of the intensity of the main peak assigned to the TM of the outer tube relatively to that of the unshifted lines. It must be pointed out that these results suggest that the band at 1548 cm$^{-1}$ should be assigned to the tangential modes ($E$ symmetry) of the inner tubes. In previous studies on p-type doped single-walled carbon nanotubes, an upshift of the TM was assigned to the charge transfer between dopant (electron acceptor) and the tube \cite{rao1,grigorian}. As a consequence, both the upshift of the TM related to the outer tubes and the non-shift of the TM related to the inner tubes states that the charge transfer only occurs between iodine ions and the carbon atoms of the outer tubes. \newline

Additional information can be found from the comparison of Raman spectra of pristine and doped DWCNT excited with laser wavelengths 647.1  nm (figure 3) and 514.5 nm (figure 4). In the pristine samples, couples of RBM assigned to inner and outer tubes are observed. In the Raman spectrum measured with the 647.1 nm excitation wavelength (figure 3, top-left), DWCNTs featured by the following couples of bands are identified: (i) 122 cm$^{-1}$ and 193 cm$^{-1}$ , (ii) broad band around 146 cm$^{-1}$ related to thin lines at 252 cm$^{-1}$, 261 cm$^{-1}$, 281 cm$^{-1}$ and 289 cm$^{-1}$, (iii) line at 171 cm$^{-1}$ associated to the line at 340 cm$^{-1}$. In the spectrum measured with the 514.5 nm excitation wavelength (figure 4, top-left), the RBM couples are the following: (i)a broad band around 150 cm$^{-1}$ is related to the thin lines at 243 cm$^{-1}$, 261 cm$^{-1}$ and 268 cm$^{-1}$, (ii) the band at 165 cm$^{-1}$ is associated to the line at 312 cm$^{-1}$. In contrast, the band at 205 cm$^{-1}$ does not display a counterpart in the high frequency range. As a consequence, it is assigned to the RBM of a single-walled carbon nanotube. In a previous Raman investigation performed on the same kind of sample, the same assignment was proposed for a line at 204 cm$^{-1}$ \cite{bacsa}.\newline 
In iodine-doped samples, the low-frequency range is dominated by peaks at 110 cm$^{-1}$ and 158 cm$^{-1}$ in the spectra measured with the 647.1 nm excitation wavelength, and by peaks located at 105 cm$^{-1}$, 160 cm$^{-1}$ and 174 cm$^{-1}$ in the spectra measured with the 514.5 nm excitation wavelength (Figures 3 and 4, bottom-left). In previous studies \cite{hsu}, these lines were assigned as the resonant modes of poly-iodine anions, mainly I$_{3}^{-}$ and I$_{5}^{-}$. These strong modes overlap with the RBM response of the outer tubes in DWCNT. However, with regards to the strong intensity of these RBM modes in pristine samples (Figure 3 and 4, top-left), it seems obvious that we should be able to observe these RBMs surperimposed to the poly-iodine modes in iodine-doped samples if they exist. The non-observation of these bands confirms the result obtained using the 1064 nm excitation wavelength. The RBMs of the outer tubes vanish upon doping. In contrast, RBMs assigned to the inner tubes are always observed in the high frequency range of the iodine-doped samples, however with a smaller intensity than in pristine DWCNT. The information on p-doped DWCNT, provided from the RBM range of the Raman spectra excited at 514.5 nm and 647.1 nm, is really new. Indeed in Br-doped DWCNT, the RBM region of the Raman spectrum excited at 514.5 nm is not actually discussed because the stretching frequency of bromine vapor at $\sim$ 320 cm$^{-1}$ overlaps with the RBM frequency of the inner tubes \cite{chen}. Here, the excitation dependence of the intensity of the RBM of the inner tubes in iodine-doped DWCNT can be discussed. In contrast with the strong intensity of the RBM assigned to the semiconducting inner tubes in the Raman spectrum excited at 1064 nm (figure 2, left), a weak intensity of the RBM is observed in the Raman spectra excited at 514.5 nm and 647.1 nm. Especially, the RBM intensity is smaller in p-doped DWCNT than in pristine DWCNT (figure 3 and 4, left). Using the single-resonance Raman approach \cite{kataura}, we found that the RBMs that appear the most affected in the Raman spectrum measured with the 514.5 nm and 647.1 nm  excitation wavelenghts can be assigned to the metallic inner tubes. These are the RBM in the 193-216 cm$^{-1}$ range in the Raman spectrum measured a 647.1 nm, and the RBM in the 243-270 cm$^{-1}$ range in the Raman spectrum measured a 514.5 nm. These results suggest a peculiar sensitivity of the metallic inner tubes upon doping. This surprising result have to be unambiguously confirmed. In this aim, it will be interesting to make experiments on DWCNTs samples in which intense RBMs related to metallic and semiconducting tubes coexist together before doping.\\ 

In the Raman spectrum of the pristine sample measured with the 647.1 nm excitation wavelength (figure 3, top-right), the high frequency part of the TM band can be fitted by a strong line at 1586 cm$^{-1}$ and a weak component at 1600 cm$^{-1}$. These lines are located at 1590 cm$^{-1}$ and 1605 cm$^{-1}$ in the Raman spectrum measured with the 514.5 nm excitation wavelength (figure 4, top-right). The low frequency part of the TM profile measured with the 647.1 nm excitation wavelength is well described by a narrow and broad component located at 1583 cm$^{-1}$ and 1550 cm$^{-1}$, respectively. In the spectrum measured with the 514.5 nm excitation wavelength , three components must be taken into account: a weak and narrow line around 1582 cm$^{-1}$ and two broad bands at 1564 cm$^{-1}$ and 1535 cm$^{-1}$. In agreement with calculations \cite{rahmani}, the two highest frequency components are assigned to $A_{1}$ and $E$ tangential modes of the outer tubes, the line around 1585 cm$^{-1}$ is the $A_{1}$ TM of the inner tube, the broad component at the lowest frequency is attributed to $E$ TM of the inner tube and that around 1564 cm$^{-1}$ to $E$ TM of the outer tube. The doping dependence of all these modes confirms these assignments. Indeed, all the bands assigned to the TMs for the outer tubes upshift upon doping, of about 6 cm$^{-1}$ in the spectrum measured with the 647.1 nm excitation wavelength, and about 10 cm$^{-1}$ in the spectrum measured with the 514.5 nm excitation wavelength. The lines assigned to the TMs for the inner tubes do not shift upon doping. As already discussed here (for a 1064 nm excitation wavelength), the relative intensities of all the shifted bands significantly decrease upon doping with respect to those of the non-shifted components. Recently the Raman spectrum of bromine doped DWCNTs was studied using a 514.5 nm excitation source. An upshift of about 16 cm$^{-1}$ of all the modes assigned to outer tubes: 1592 cm$^{-1}$, 1601 cm$^{-1}$, 1573 cm$^{-1}$, and 1552 cm$^{-1}$ lines, simultaneously with a decrease of their intensities with respect to that of the non-shifted mode at 1580 cm$^{-1}$ were clearly evidenced (figure 2 of Ref.\cite{chen}). These results are in agreement with this present study. All these results unambiguously state that iodine doping mainly affects the outer tube vibrations. Thus, the adsorption of iodine species only occurs on the surface of the outer tubes. Nevertheless, a peculiar sensitivity upon doping of the metallic inner tube is suggested. Its origin is an open question. All the previous results have been analyzed in a consistent way in a single-resonance approach. However, it was recently suggested that a double-resonance process could describe the behavior of the Raman spectrum with the excitation energy in single-wall carbon nanotubes, especially the multiple-peak structure in the TM range \cite{maultzsch1,maultzsch2}. Obviously, this approach cannot be ruled out from the present results. Nevertheless, we think that its implication does not alter our main conclusion. Indeed, the iodine adsorption at the surface of the outer tubes which is revealed by this study is an experimental fact, independently of the approach used to discuss the data. Only the excitation dependence of the shift of the TM with the incident wavelength could be discussed more precisely in terms of a double-resonance process. \newline

\section{conclusion}

In summary, we have performed a detailed Raman study of pristine DWCNTs samples grown by a CCVD method and of the same tubes doped with iodine. For the pristine DWCNTs, all RBMs were assigned to inner and outer tube couples. These assignments were obtained from calculations based on the bond polarization theory, using the spectral moment method. Some lines which did not have a counterpart were assigned to single-walled carbon nanotubes. The Raman investigation of the iodine-doped DWCNT samples were performed using three different wavelengths. The results obtained state that poly-iodine anions are adsorbed on the surface of the outer tubes of the DWCNTs, and that the charge transfer only occurs between iodine ions and the carbon atoms of the outer tubes. The results and the conclusions given here are in perfect accordance with a previous investigation on the dependence of Raman spectra of DWCNTs upon bromine doping \cite{chen}. An universal behaviour of DWCNT upon a p-type doping emerges from these Raman studies.

\vskip 7mm

\newpage

\begin{figure}
\begin{center}
\includegraphics{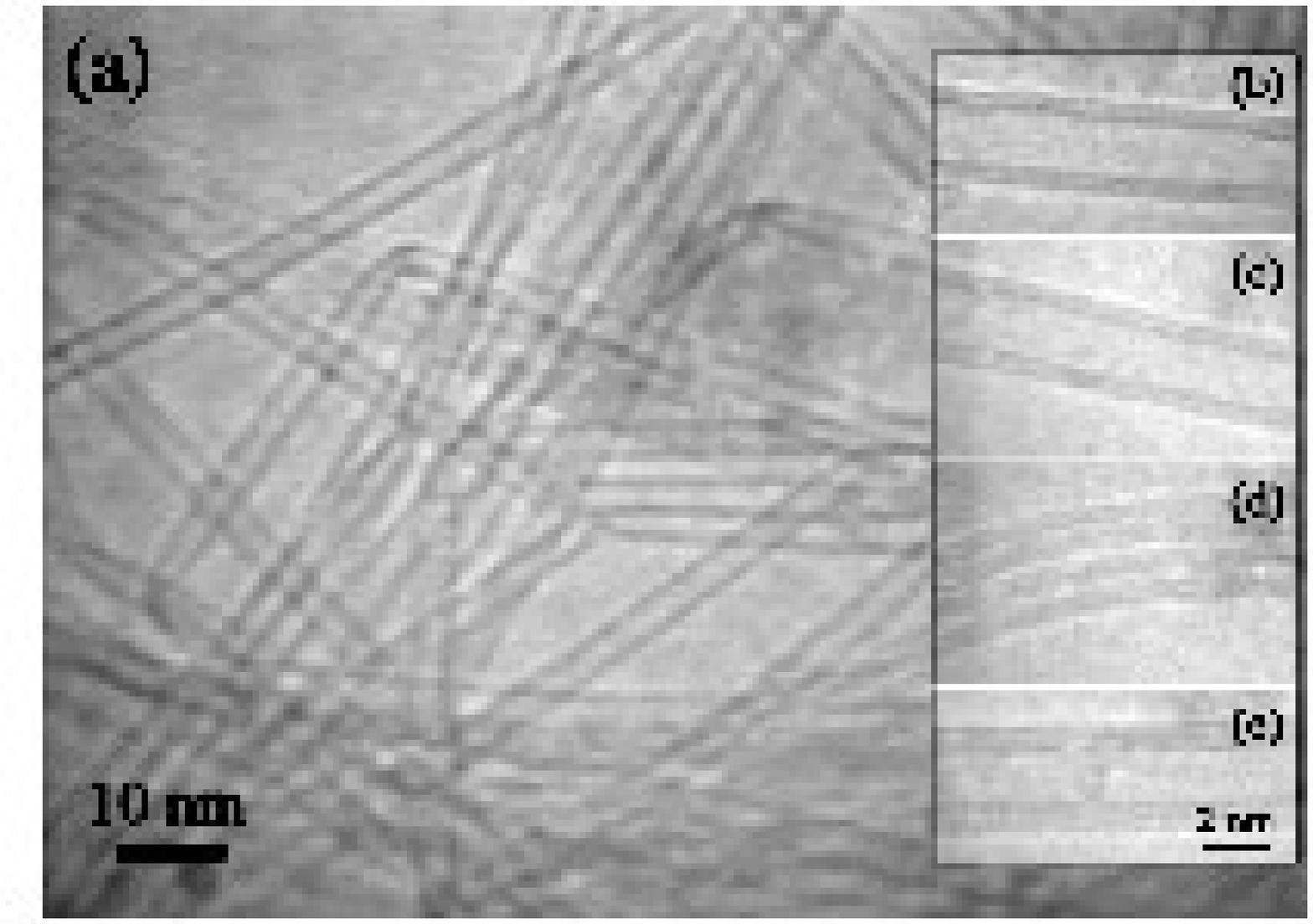}
\caption{(a) TEM image of the sample of DWCNTs. (b-d) High-Resolution TEM images of DWCNTs; (e) High-Resolution TEM image of a triple-walled CNT.}\vskip 7mm
\end{center}
\end{figure}

\begin{figure}
\begin{center}
\includegraphics{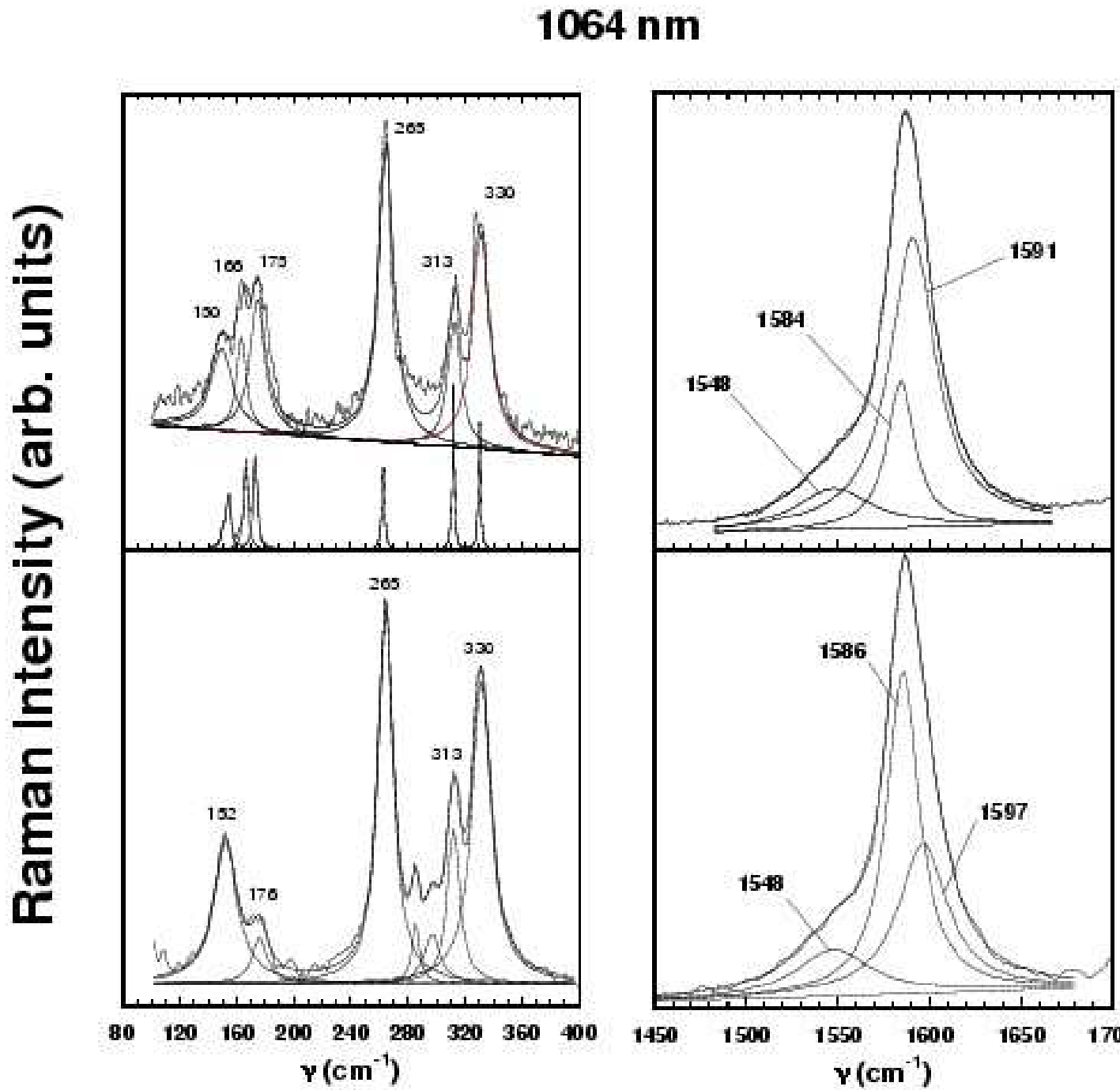}
\caption{Raman spectra measured using a 1064 nm excitation wavelength. The spectra are fitted by a sum of lorentzians. Top: pristine DWCNT sample; left: RBM range, (bottom solid line: calculated Raman spectra, see text); right: TM range. Bottom: iodine-doped DWCNT sample; left: RBM range, right: TM range}\vskip 7mm
\end{center}
\end{figure}

\begin{figure}
\begin{center}
\includegraphics{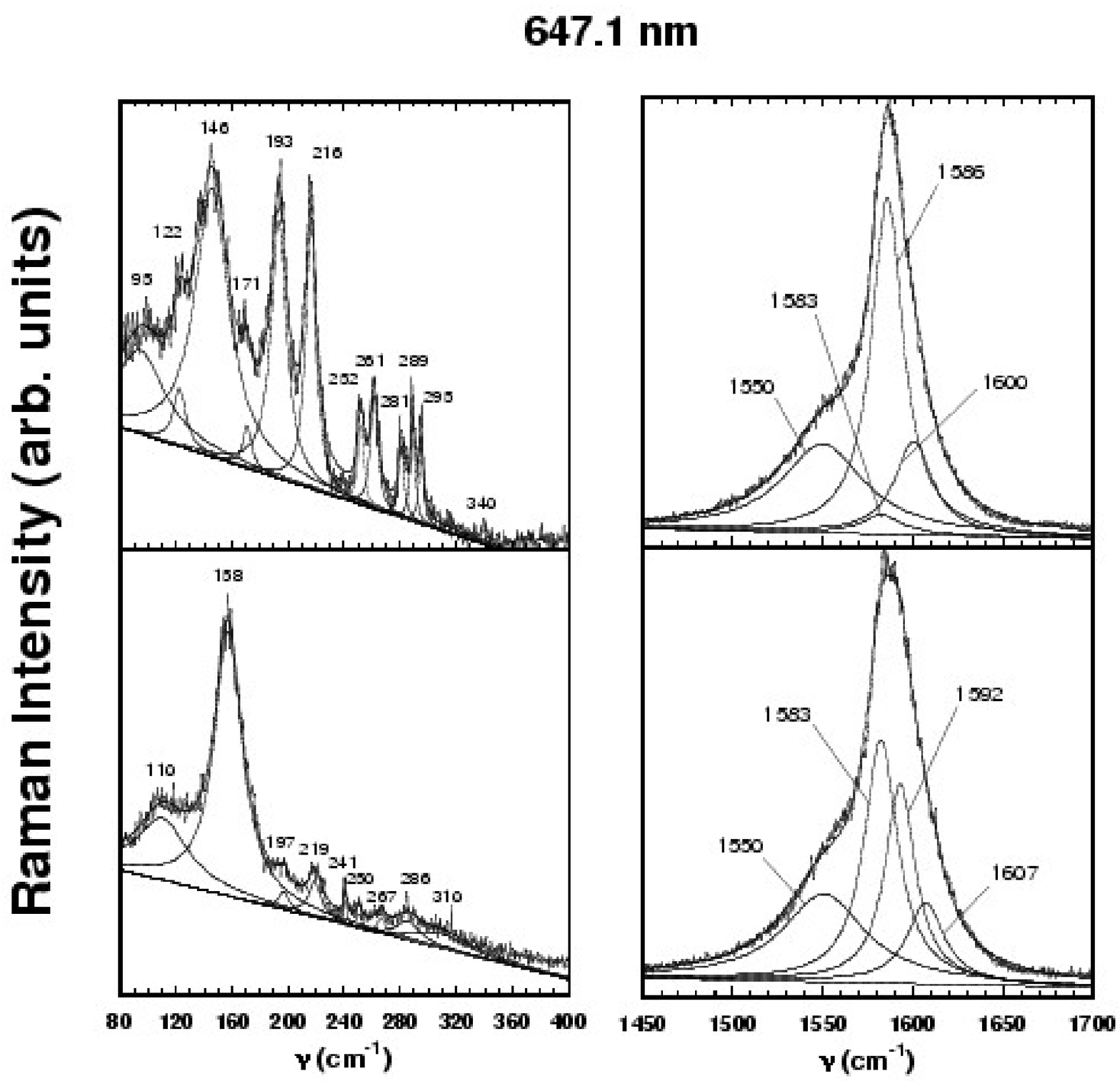}
\caption{Raman spectra measured using a 647.1 nm excitation wavelength. The spectra are fitted by a sum of lorentzians. Top: pristine DWCNT sample; left: RBM range, right: TM range. Bottom: iodine-doped DWCNT sample; left: RBM range, right: TM range}\vskip 7mm
\end{center}
\end{figure}

\begin{figure}
\begin{center}
\includegraphics{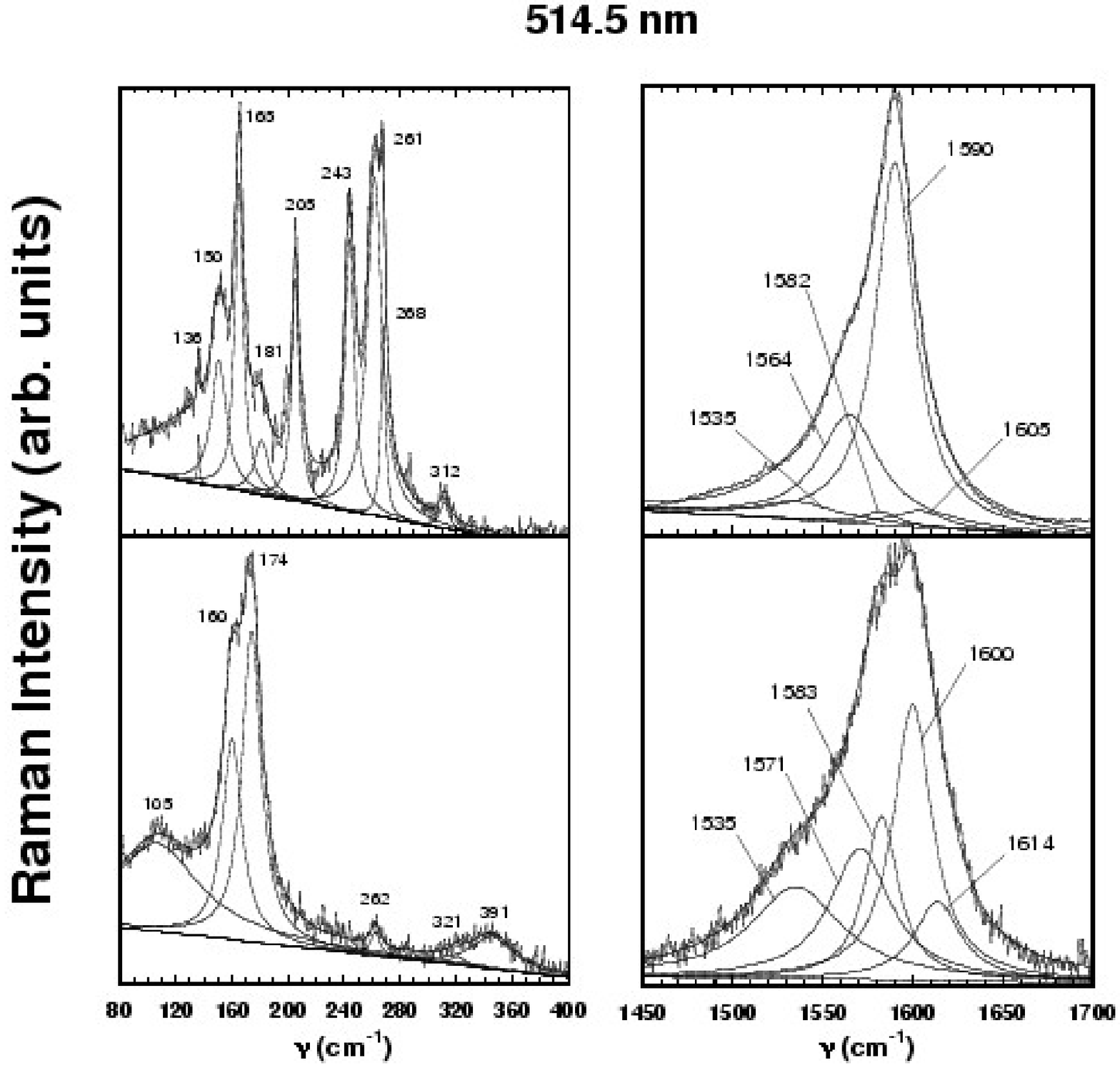}
\caption{Raman spectra measured using a 514.5 nm excitation wavelength. The spectra are fitted by a sum of lorentzians. Top: pristine DWCNT sample; left: RBM range, right: TM range. Bottom: iodine-doped DWCNT sample; left: RBM range; right: TM range}\vskip 7mm
\end{center}
\end{figure}

\end{document}